\documentclass[runningheads]{llncs}
\usepackage[T1]{fontenc}
\usepackage{cite}
\usepackage{graphicx}
\usepackage{multirow}
\usepackage{amsmath}
\usepackage{makecell}
\usepackage{hyperref}
\usepackage{enumitem}
\usepackage{multirow}
\usepackage{boldline}
\usepackage{hhline}
\usepackage{CJKutf8}
\usepackage{float}
\usepackage{xcolor}
\usepackage{marvosym}

\usepackage{caption}
\usepackage{tabularray}
\begin{document}
\title{Enhancing Depression-Diagnosis-Oriented Chat with Psychological State Tracking}
\titlerunning{Depression-Diagnosis-Oriented Chat with Psychological State Tracking}
%
\author{
\textbf{Yiyang Gu}$^{1\star}$, \textbf{Yougen Zhou}$^{2}$\thanks{Equal contribution.}, \textbf{Qin Chen}$^{1}$\textsuperscript{(\Letter)}, \\
\textbf{Ningning Zhou}$^3$, \textbf{Jie Zhou}$^{1}$, \textbf{Aimin Zhou}$^{2}$, \textbf{Liang He}$^{1,2}$ 
}
\authorrunning{Y. Gu et al.}
%
\institute{
School of Computer Science and Technology, East China Normal University \and
Shanghai Institute of Al for Education, East China Normal University \and
School of Psychology and Cognitive Science, East China Normal University \\
\email{\{yygu,zyg\}@stu.ecnu.edu.cn, qchen@cs.ecnu.edu.cn}}
%
\maketitle              
\begin{abstract}
Depression-diagnosis-oriented chat aims to guide patients in self-expression to collect key symptoms for depression detection. Recent work focuses on combining task-oriented dialogue and chitchat to simulate the interview-based depression diagnosis. However, these methods can not well capture the changing information, feelings, or symptoms of the patient during dialogues. Moreover, no explicit framework has been explored to guide the dialogue, resulting in some ineffective communications that impact the experience.
In this paper, we propose to integrate \textbf{P}sych\textbf{o}logical \textbf{S}tate \textbf{T}racking (POST) within the large language model (LLM) to explicitly guide depression-diagnosis-oriented chat. Specifically, the state is adapted from a psychological theoretical model, which consists of four components: Stage, Information, Summary, and Next. We fine-tune an LLM model to generate the dynamic psychological state, which is further used to assist response generation at each turn to simulate the psychiatrist. Experimental results on the existing benchmark show that our proposed method boosts the performance of all subtasks in depression-diagnosis-oriented chat. 

\end{abstract}
\section{Introduction}
Depression remains an escalating mental health threat globally, due to the severe scarcity and limited access to professionals. To alleviate such situations, conversational agents become a promising solution for early depression detection, due to the traditional detection mechanisms being invasive \cite{chaytor2003ecological}. In practical psychotherapy, psychiatrists dynamically adjust the dialogue flow to effectively collect symptoms from patients, while providing appropriate intervention strategies such as emotional support. To simulate this process, Yao et al. \cite{yao-etal-2022-d4} defined this kind of dialogue as \emph{Task-oriented Chat} and collected the first dialogue dataset D$^4$ for depression diagnosis. 
However, existing work mostly focuses on shallow heuristic attempts such as predicting topics and generating empathetic responses, falling short of capturing the changing information, feelings or symptoms of the patient during dialogues.


Recently, Large Language Models (LLMs) have achieved remarkable success in various text reasoning tasks. In the field of psychology, ChatGPT 
\cite{schulman2022chatgpt} and GPT-4 
\cite{openai2023gpt4} have shown promising performance in attributing mental states 
\cite{bubeck2023sparks}. While researchers have envisioned further harnessing this capability for complex psychological tasks, the majority still focus on developing chatbots for emotional support purposes. Moreover, no explicit framework has been explored to guide the dialogue, creating a gap between LLMs and psychiatrists, especially regarding their lack of skill in questioning 
\cite{chiu2024computationalframeworkbehavioralassessment}.

To enhance depression-diagnosis-oriented chat, we propose the \textbf{P}sych\textbf{o}logical \textbf{S}tate \textbf{T}racking (POST) to link the patient’s current symptom with the doctor’s next strategy. Inspired by Albert Ellis’ ABC Model \cite{ABC} in Cognitive-behavioral therapy (CBT), we define the psychological state with four components: \emph{Stage}, \emph{Information}, \emph{Summary} and \emph{Next}. First, we figure out at which stage of the current diagnosis procedure is; Then, we distinguish the key symptoms information that the patient is exhibiting; After that, we document the current diagnostic summary of the patient; Finally, in \emph{Next}, we introduce a targeted prompt to align with specific counseling strategies. We jointly optimize the POST model and the response generation model by an LLM. Experimental results on the existing benchmark show that our proposed method achieves the best performance of all subtasks in depression-diagnosis-oriented chat. Furthermore, psychological state tracking, as the explicit thought behind response generation, provides professional-compliant interpretability to the diagnostic process.
The main contributions of our work are as follows:
\begin{enumerate}
    \item[$\bullet$] We annotate a fine-grained dataset by augmenting the D$^4$ dataset, which annotates the psychological state of each conversation round guided by the ABC Model.
    \item[$\bullet$] We propose a joint model to explicitly guide depression-diagnosis-oriented chat, which integrates psychological state tracking into an LLM to learn the connection between patients' state changes and doctors' strategic planning.
    \item[$\bullet$] Extensive experiments on the existing benchmark show that our proposed method boosts the performance of depression-diagnosis-oriented chat. In particular, the psychological state tracking serves as an explicit thought to provide interpretability for response generation.
\end{enumerate}

\section{Related Work}
\subsection{Depression Diagnosis}
Depression diagnosis aims to use diagnostic tools to identify symptoms and determine the severity of depression \cite{compas1993taxonomy,mitchell2009clinical}. Early research was conducted by psychiatrists in controlled settings through self-questionnaires, such as the PHQ-9 \cite{kroenke2001phq} and GAD-7 \cite{spitzer2006brief}, to assess patients' cognitive or emotional states. However, in face-to-face settings, individuals often hesitate to express their mental state. Some researchers explore using different network structures to automatically identify mental health status in social media content \cite{bucur2023s}. The poor interactivity of these approaches also limits patients' self-expression. Yao et al. \cite{yao-etal-2022-d4} proposed to combine task-oriented dialogue and chitchat to simulate the interview-based depression diagnosis. Seo et al. \cite{seo2024diagescdialoguesynthesisintegrating} integrated depression diagnosis into emotional support conversation to improve diagnosis ability. Whereas, these methods struggle to well capture the changing information, feelings, or symptoms of the patient during the dialogue process. Moreover, there has been no exploration of an explicit framework to guide the response generation. Thus, we aim to explore a more personalized and professional depression diagnostic chatbot.



\subsection{Dialogue State Tracking}

Dialogue State Tracking (DST) \cite{williams2016dialog} is essential in task-oriented dialogue systems for monitoring conversation states. Previous studies have utilized pre-trained models to improve DST. For example, Wu et al. \cite{wu2023semanticparsinglargelanguage} employed large-scale pre-trained models for zero-shot DST and enhanced dialogue state tracking with intricate updating strategies. Sun et al. \cite{sun_tracking_2022} revolutionize dialogue state tracking with a Mentioned Slot Pool (MSP) to improve accuracy. However, these approaches cannot effectively handle complex dialogues and capture fine-grained semantic relationships.

The rise of large language models has led to advancements in DST methods. Recent research focuses on prompt learning \cite{yang_dual_2023}, meta-learning \cite{chen_stabilized_2023}, and LLM agents \cite{niu2024enhancingdialoguestatetracking} to enhance DST performance. Despite these progressions, there remains a gap when compared to more advanced models like ChatGPT, and additional support from psychological theories is necessary to accomplish our task. Therefore, we designed the DST framework based on the ABC model of Cognitive-Behavioral Therapy and integrated it into large language models (LLMs) to guide the generation of responses for depression diagnosis. Fine-tuning techniques such as LoRA \cite{lora} are also applied to further improve the generation performance of the LLMs.

\section{Data Annotation}

\subsection{Annotation Procedure}
We adapt the D$^{4}$ \cite{yao-etal-2022-d4} dataset with additional psychological state annotations, which contains 1,339 clinically standardized conversations about depression. The clinical data can facilitate a generation and diagnosis process that closely simulates real-life clinical consultations for depression. However, the original data lacked tracking of the patient's conditions. 

\begin{figure}[h]
    \centering
    \includegraphics[scale=0.55]{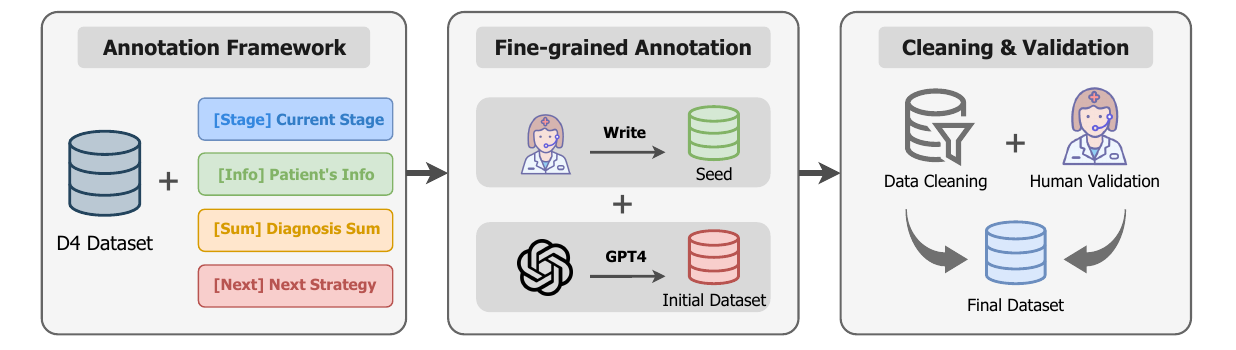}
    \caption{Annotation Procedure}
    \label{fig:annotation_process}
\end{figure}

To transform the raw data into a sample dataset that can be used for psychological state tracking, we annotate the conversations following three steps, as shown in Figure \ref{fig:annotation_process}: 
\begin{enumerate}
    \item[$\bullet$]First, we constructed an annotation framework to capture psychological states in real-time during the conversations. Special tokens were added to the beginning of each utterance to indicate the Current \textbf{Stage} based on Albert Ellis' ABC Model, Patient's \textbf{Info}rmation, Diagnosis \textbf{Sum}maries, and \textbf{Next} Strategies. The current stage refers to the three key stages in ABC Model: Stage-A focuses on identifying \textbf{A}ctivating events that trigger emotional responses; Stage-B centers on understanding the patient's \textbf{B}eliefs and thought patterns regarding these events; and Stage-C evaluates the emotional and behavioral \textbf{C}onsequences resulting from these beliefs.
    \item[$\bullet$]Second, we implemented fine-grained data annotation using a system developed based on LabelStudio \footnote{\href{https://labelstud.io}{https://labelstud.io}}. Three professional psychologists with expertise in clinical depression consultations were invited to perform the initial manual annotations following standard clinical protocols. These carefully annotated samples then served as seeds for GPT-4 to extend the annotation process to the remaining dataset, balancing annotation quality with efficiency while maintaining professional standards throughout the process.
    \item[$\bullet$]Third, we conducted thorough data cleaning and validation on the annotated dataset. To ensure the quality and reliability of GPT-4 annotations, our psychologists manually reviewed and verified all labels generated by GPT-4. 
\end{enumerate}


\subsection{Data Analysis}



\begin{table}[h]
    \centering
    \begin{minipage}[c]{0.48\textwidth}
        \captionof{table}{Statistics of annotated D$^4$}
        \centering
        \resizebox{1\textwidth}{!}{
        \begin{tabular}{clc}
            \hlineB{4}
            \textbf{Source} & \textbf{Criteria} & \textbf{Total}\\
            \hline
            \multirow{7}{*}{\makecell{D$^4$}} & Dialogues & 1,339 \\
            & Dialogue turns & 28,977 \\
            & Average turns per dialogue & 21.67 \\
            & Average tokens per dialogue & 577.12 \\
            & Average tokens per utterance & 13.31 \\
            & Average patient tokens per utterance & 11.87 \\
            & Average doctor tokens per utterance & 14.76 \\
            \hline
            \multirow{6}{*}{POST} & Stage-A per dialogue & 5.84 \\
            & Stage-B per dialogue & 4.11 \\
            & Stage-C per dialogue & 11.72 \\
            & Average Info tokens per turn & 47.12 \\
            & Average Summary tokens per turn & 22.69 \\
            & Average POST tokens per turn & 75.81 \\
            \hlineB{4}
        \end{tabular}
        }
        \label{tab:data-statistic}
    \end{minipage}
    \hfill
    \begin{minipage}[c]{0.5\textwidth}
        \centering
        \includegraphics[width=\linewidth]{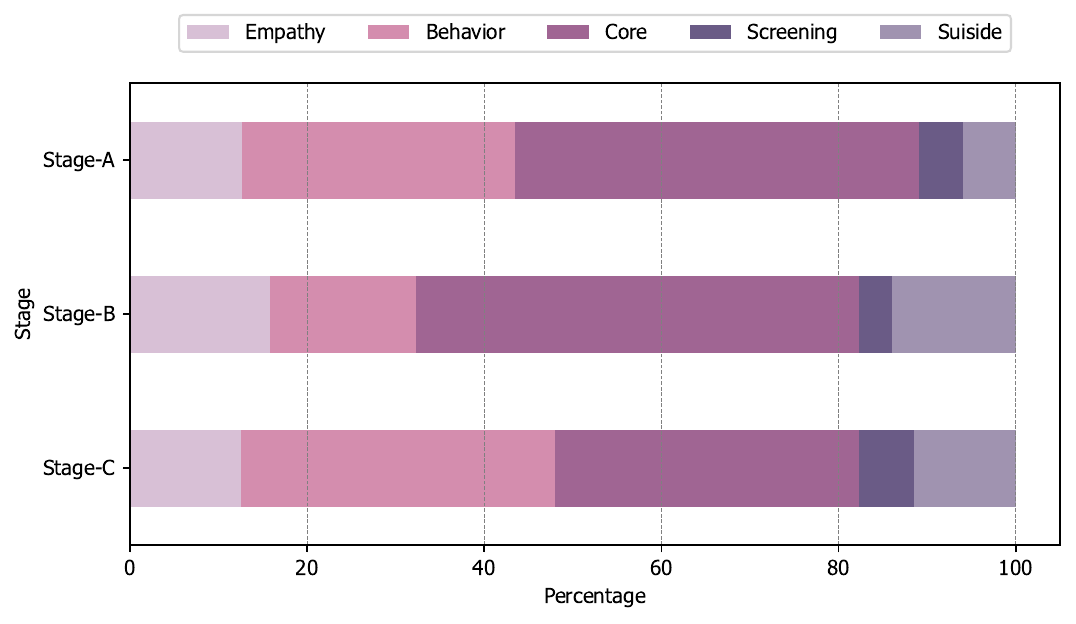}
        \captionof{figure}{Distribution of strategies in different stages}
        \label{fig:stage_dis}
    \end{minipage}
\end{table}

\subsubsection{Statistics}
The basic statistics of the annotated dataset are shown in Table \ref{tab:data-statistic}. The total dialogues in the dataset are 1,339, while the total turns of dialogues are 28,977. Due to the data cleaning procedure, the remaining dialogue turns may be less than the number in the original D$^4$ dataset. The average number of doctor tokens per utterance is 14.76, which is approximately 3 tokens more than the average number of patient tokens, indicating that doctors often speak more due to the need for consultation or empathetic consolation. The second part concerns the statistical analysis of annotated POSTs. The average number of POST tokens per turn is 75.81, suggesting that POSTs contain more information compared to utterances. Notably, the Info part, which contains patient information from dialogue history, is lengthier with an average length of 47.12. In contrast, the Summary, which involves further inference of the patient's diagnostic results, is almost half the length of the Info part, averaging only 22.69.

\subsubsection{Stage Analysis}
The distribution of the next strategies in different stages is illustrated in Figure \ref{fig:stage_dis}. The chart reveals that Stage-A primarily focuses on Core and Behavior, aligning with its objective of identifying triggering events. In Stage-B, there is a notable increase in attention to Empathetic Comfort and Suicidal tendencies. Conversely, Stage-C, as the terminal phase, primarily focuses on evaluating behavioral outcomes and intensifies screening to facilitate final diagnoses. 




\section{Method}
\begin{figure*}[h]
    \centering
    \includegraphics[scale=0.60]{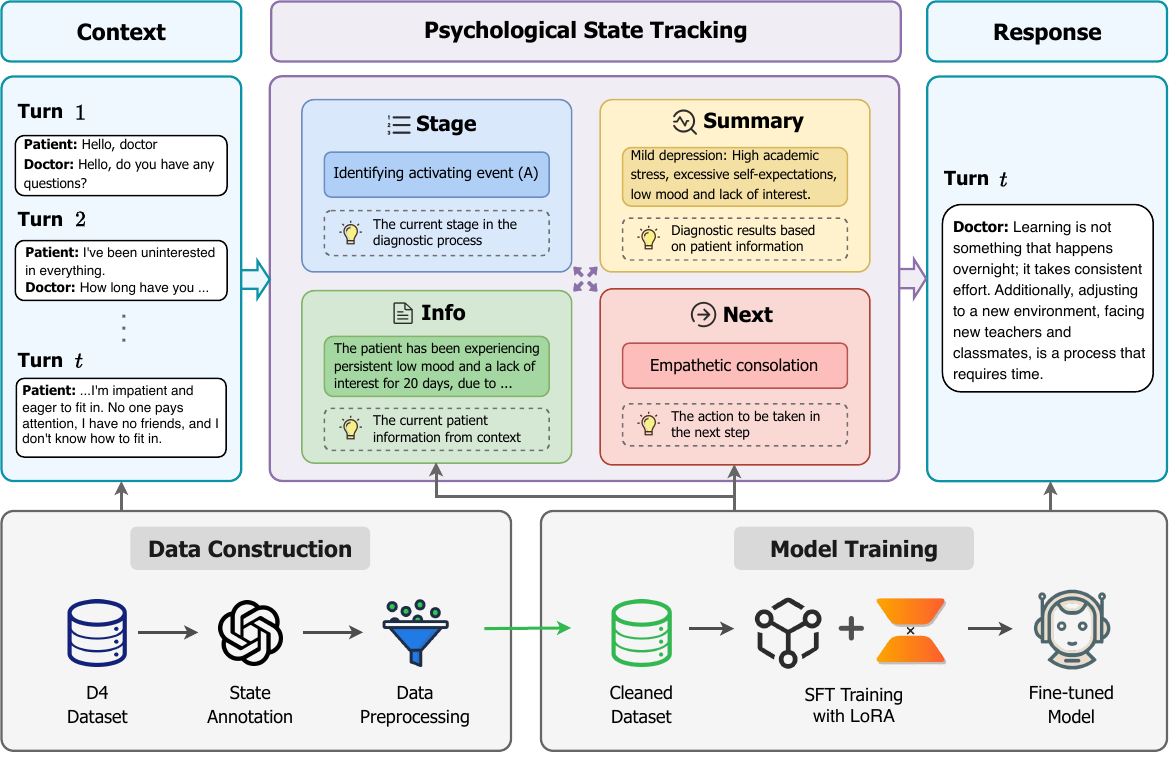}
    \caption{The overall framework of depression-diagnosis-oriented chat with psychological state tracking.}
    \label{fig:framework}
\end{figure*}

Our approach formalizes depression-diagnosis-oriented chat by representing the user's psychological state as a set of task attributes. The goal is to generate doctors’ probable responses based on the dialog context, taking into account the current state and next planning. As shown in Figure \ref{fig:framework}, following the task-specific fine-tuning paradigm, we build a joint model for psychological state tracking and response generation by equipping a transformer-based language backbone with functional modules.

\subsection{Task Formulation}
For a depression-diagnosis-oriented chat task, there is a $t$ turn dialogue between a patient and a doctor that can be represented as:
\begin{equation}
    d_t = (u_1^{p}, u_1^{d}, u_2^{p}, u_2^{d}, \cdots, u_{t}^{p},  u_{t}^{d})
\end{equation}
where $u_t^p$ is the patient's utterance, and $u_t^d$ is the doctor's response at turn $t$. The entire depression-diagnosis-oriented chat procedure can be split into 2 subtasks: \textbf{Psychological State Tracking} predicts the patient's current psychological state based on the dialogue context, which includes \textit{Stage}, \textit{Information}, \textit{Summary} and \textit{Next}. \textbf{Response Generation} generates the most likely response based on the dialogue history and current state. We jointly optimize the psychological state tracking model and the response generation model by an LLM. 

\subsection{Psychological State Tracking}
Clinically, interview-based depression diagnosis needs to collect and summarize key symptom information about the patient while providing a chat-like conversation experience. Psychiatrists need to design the questioning logic between questions of symptoms from mild to severe during the consultation. To model such fine-grained relationships between patients' symptoms and questions, we perform psychological state tracking. The state consists of the following components:
\begin{equation}
    State_t = [S_t; I_t; Sum_t; N_t]
\end{equation}
where $S_t, I_t, Sum_t, N_t$ stand for Stage, Information, Summary and Next of dialogue turn $t$.

\paragraph{Stage}
In clinical practice, a consultation follows a gradual in-depth manner and diagnosis strategies consistently occur across turns. For example, doctors usually start by asking about core symptoms such as mood and interests, and then gradually turn to behavior symptoms. To perform a deep analysis of the strategy transition, we first need to find out at which stage of the current diagnosis procedure. After an assessment, we summarize the dialogue stage into Stage-A, Stage-B, or Stage-C, which represents Identifying \textbf{A}ctivating Events, Perceiving Patients' \textbf{B}eliefs, and Assessing \textbf{C}onsequences respectively. The process can be formulated as:
\begin{equation}
    S_t = \pi_{\text{POST}}(h_t;\theta) \\
\end{equation}
where $S_{t-1}$ is the stage of previous turn, $\pi_{\text{POST}}$ represents our POST model, $\theta$ denotes the parameters of the model, and $h_t$ represents the current dialogue history, which consists of $t-1$ turns of dialogue and patient's utterance at turn $t$:
\begin{equation}
    h_t = (u_1^{p}, u_1^{d}, u_2^{p}, u_2^{d}, \cdots, u_{t-1}^{p},  u_{t-1}^{d}, u_t^p) \\
\end{equation}

\paragraph{Information}
This part aims to discover the psychological information that exhibited in the dialogue history. Based on the presented symptoms, we derive the patient's illness severity. For healthy individuals, the conversation typically manifests surface symptoms such as changes in sleep. As the condition worsens, patients tend to exhibit an increasing array of symptoms. Then we record the symptoms info by the following formulation:
\begin{equation}
    I_t = \pi_{\text{POST}}(h_t,S_t;\theta) \\
\end{equation}
where $I_t$ represents the information of dialogue at turn $t$. By documenting the symptoms exhibited by the patient as a form of memory, we can gain a clearer understanding of their depression status and design strategies for responses. 

\paragraph{Summary}
This subsequence aims to document the current diagnosis summary of the patient. Depression diagnoses are primarily employed for preliminary screening. For milder cases, empathetic strategies are generally used to encourage the patient's self-expression. But for severe cases, immediate crisis intervention is required. Therefore, we have incorporated real-time diagnostic results into the process of depression diagnosis, formulated as:
\begin{equation}
    Sum_t = \pi_{\text{POST}}(h_t,S_t,I_t;\theta) \\
\end{equation}

\paragraph{Next}
To facilitate dialogue generation, we introduce the \textit{Next} strategy to guide the LLM, which considers the current stage, the patient's symptom information, and the severity of depression. It determines which strategy and topic should be used in the next response to support further diagnosis:
\begin{equation}
    N_t = \pi_{\text{POST}}(h_t,S_t,I_t,Sum_t;\theta) \\
\end{equation}

After the LLM finishes the generation for all psychological states, we prompt it with a combined pair of the states and the current dialogue history to generate the doctor's response:
\begin{equation}
    u_t^d = \pi_{\text{POST}}([h_t; State_t]; \theta)
\end{equation}
With psychological state tracking, we obtain a fully interpretable thought process for generating responses focused on depression diagnosis. 

\subsection{Fine-tuning}
In this work, we fine-tune large language models with a parameter-efficient approach, i.e., Low-Rank Adaptation (LoRA). LoRA maintains the weights of pre-trained LMs while introducing trainable rank decomposition matrices into each transformer layer, making it feasible to fine-tune LLMs with much fewer computational resources.

We fine-tune an LLM to track the psychological state and generate the response jointly, given the crafted example and the annotated label. Specifically, the objective is to predict the next token based on language modeling:
\begin{equation}
    \min \limits_{\theta} \sum_{t=1}^{T} - {\log} {p}_{\theta} (State_t, u_t^d|h_{t}; \theta)
\end{equation}
where $\theta$ represents the parameters for a language model and $T$ is the total turns of the dialogue. Ideally, the objective encourages the model to learn the target distribution by predicting tokens in the sequence. By placing the psychological state before the doctor's response, the model learns to fuse the distribution from thought to response in an in-context language modeling manner. We only compute the loss of tokens on the psychological state $State$ and the doctor utterance $u^d$.

\section{Experiments}

\subsection{Experimental Setups}
\subsubsection{Baselines}
We leverage the CPT model \cite{cpt} as our primary baseline, as it achieved the best performance in previous studies \cite{yao-etal-2022-d4}. And we use the same configuration as them. ChatGPT (gpt-3.5-turbo) \footnote{\href{https://openai.com/blog/chatgpt}{https://openai.com/blog/chatgpt}} is also listed as a baseline, which performs well in most tasks.

\subsubsection{Implementation Details}
Our model uses ChatGLM3-6b\footnote{\href{https://github.com/THUDM/ChatGLM3}{https://github.com/THUDM/ChatGLM3}} as the base architecture, and is implemented in PyTorch. During the supervised fine-tuning process, we apply LoRA to all linear layers of the model, where LoRA rank is set to 64. We set the batch size, max context length, and learning rate to 32, 1024, and 2e-4, respectively. The model is trained on one A800 GPU for 5 epochs, which costs about 8 hours. 

\begin{table}[h]
\caption{Results of automatic evaluation. ``*'' means golden labels are given.}
\centering
\resizebox{1\textwidth}{!}{
\setlength{\tabcolsep}{1.5pt}{ 
\begin{tabular}{llcccccc}
\hlineB{4}
\textbf{Model} & \textbf{Settings} & \textbf{BLEU-2} & \textbf{ROUGE-L} & \textbf{METEOR} & \textbf{DIST-2} & \textbf{Next ACC.} \\
\hline
ChatGPT    & /            & 10.01\%     & 0.19        & 0.3254     & 0.15       & -              \\
CPT        & /            & 19.79\%     & 0.36        & 0.2969     & 0.07       & -              \\
ChatGLM3   & /            & \textbf{30.85\%} & \textbf{0.47} & \textbf{0.4524} & \textbf{0.17}       & -              \\
\hline
ChatGPT    & +POST       & 17.52\%     & 0.30        & 0.4080     & \textbf{0.19}   & 14.62\%        \\
           & $\mathrm{\Delta}$ & \textcolor{blue}{+7.51\%} & \textcolor{blue}{+0.11} & \textcolor{blue}{+0.0826} & \textcolor{blue}{+0.04} & -  \\
CPT        & +POST       & 34.15\%     & 0.44        & 0.4609     & 0.06       & 57.12\%        \\
           & $\mathrm{\Delta}$ & \textcolor{blue}{+14.36\%} & \textcolor{blue}{+0.08} & \textcolor{blue}{+0.1640} & \textcolor{red}{-0.01} & - \\
ChatGLM3   & +POST       & \textbf{39.76\%} & 0.50 & \textbf{0.5305} & 0.11       & 56.98\%        \\
           & $\mathrm{\Delta}$ & \textcolor{blue}{+8.91\%} & \textcolor{blue}{+0.03} & \textcolor{blue}{+0.0781} & \textcolor{red}{-0.06} & - \\
\hline
ChatGPT    & +POST*     & 29.46\%     & 0.42        & 0.5158     & \textbf{0.23}   & -              \\
           & $\mathrm{\Delta}$ & \textcolor{blue}{+19.45\%} & \textcolor{blue}{+0.23} & \textcolor{blue}{+0.1904} & \textcolor{blue}{+0.08} & - \\
CPT        & +POST*     & 41.94\%     & 0.55        & 0.5285     & 0.04       & -              \\
           & $\mathrm{\Delta}$ & \textcolor{blue}{+22.15\%} & \textcolor{blue}{+0.19} & \textcolor{blue}{+0.2316} & \textcolor{red}{-0.03} & - \\
ChatGLM3   & +POST*     & \textbf{45.28\%} & \textbf{0.56} & \textbf{0.5794} & 0.10       & -         \\
           & $\mathrm{\Delta}$ & \textcolor{blue}{+14.43\%} & \textcolor{blue}{+0.09} & \textcolor{blue}{+0.1270} & \textcolor{red}{-0.07} & - \\
\hlineB{4}
\end{tabular}
}}
\label{tab:generation-results}
\end{table}

\subsection{Automatic Evaluation}
We evaluate the performance of turn-based response generation given the dialogue history in the dataset.
Typical automatic metrics for text generation like BLEU-2 \cite{bleu}, Rouge-L \cite{rouge} and METEOR \cite{meteor} are employed to assess the response generation quality. In addition, we calculate DIST-2 \cite{dist} to demonstrate the diversity of generated responses. 


Table \ref{tab:generation-results} shows the results of the automatic evaluation for the response generation task. We have the following observations. 
\textbf{First}, all models show substantial improvements in BLEU-2, ROUGE-L, and METEOR after applying POST, indicating that this method can significantly enhance the generation quality.
\textbf{Second}, the application of POST resulted in considerable performance boosts across all models, with CPT showing the most significant improvement of +14.36\% in BLEU-2 over its baseline. This enhancement is attributed to the framework's ability to effectively assess the diagnostic stage, summarize patient information, and infer potential diagnoses. Further improvements are observed when golden POSTs are given.
\textbf{Third}, as ChatGPT cannot be fine-tuned for our specific task, its baseline performance in this domain is restricted. However, the introduction of POST notably improved ChatGPT's performance by +7.51\% in BLEU-2, with substantial improvements upon using golden POST (+19.45\%), demonstrating the effectiveness of the POST strategy. Interestingly, while ChatGPT shows improved DIST-2 with POST (+0.04), both CPT and ChatGLM3 exhibit slight decreases in DIST-2 (-0.01 and -0.06 respectively). This divergence reflects that unguided baselines naturally produce more diverse outputs, while POST's structure constrains variety. ChatGPT's unique characteristics allow it to maintain diversity even with structured guidance.

\subsection{Human Evaluation}
To simulate realistic depression diagnosis scenarios for evaluation, we prompt ChatGPT to act as patients, based on patient backgrounds from the dataset. Correspondingly, our model and baselines played the role of doctors, conducting diagnosis conversations with the patients. Each conversation consists of a minimum of 15 turns and will terminate at an appropriate round. Then, we assigned annotators with dialogue pairs to evaluate the performance of the doctor model following four aspects: 1) \textit{Fluency} (\textbf{Flu.}) assesses the smoothness of the whole conversation; 2) \textit{Comforting} (\textbf{Com.}) measures the ability to empathize and comfort; 3) \textit{Doctor-likeness} (\textbf{Doc.}) gauges the adaptability in shifting topics based on the patient's situation; 4) \textit{Engagingness} (\textbf{Eng.}) measures if the model sustains attention throughout the conversation. 

\begin{table}[h]
\caption{Results of human evaluation}
\centering
\setlength{\tabcolsep}{8.5pt}{ 
\begin{tabular}{ccccc}
\hlineB{4}
\textbf{Comparisons} & \textbf{Aspect} & \textbf{Win} & \textbf{Lose} & \textbf{Tie} \\ 
\hline
\multirow{4}{*}{\parbox[c]{3cm}{ChatGLM3 +POST \\ vs. \\ ChatGLM3}} & Flu. & \textbf{63.4} & 32.5 & 4.1 \\ 
                              & Com. & \textbf{95.9} & 4.1 & 0.0 \\ 
                              & Doc. & \textbf{92.7} & 6.5 & 0.8 \\ 
                              & Eng. & \textbf{93.5} & 5.7 & 0.8 \\ 
\hlineB{4}
\end{tabular}}
\label{tab:human-evaluation}
\end{table}

As shown in Table \ref{tab:human-evaluation}, the baseline using the POST method performs better in most aspects, with particularly significant improvements in \textit{Comforting}, \textit{Doctor-likeness}, and \textit{Engagingness}. Regarding fluency, the gap between our two models is not as large as other metrics, since the method has a relatively minor impact on fluency. These results suggest that incorporating the POST enhances our model's ability to emulate doctors' interactions, capturing the patient's state in real-time, and allowing for flexible strategy transitions throughout the dialogue.

\subsection{Ablation Studies}
To verify the effectiveness of our method, we conducted ablation studies. We used the model incorporating the golden POST as the baseline. Then, we removed each component of the POST in turn to study the effect of each part of the POST. The obtained results are demonstrated in Table \ref{tab:ablation-results}.

The study reveals that omitting the next strategy (\textbf{w/o Next}) significantly impacts response generation, as it is directly correlated with the subsequent response's relevance and effectiveness. Furthermore, excluding the stage (\textbf{w/o Stage}) notably affects the quality of generation, indicating the importance of the diagnostic dialogue phase in tailoring responses to be stage-appropriate and rational. In contrast, the absence of information (\textbf{w/o Info}) and summary (\textbf{w/o Sum}) components showed less impact on response generation quality, due to their roles in summarizing dialogue history rather than directly influencing responses. Nonetheless, these elements are crucial for their interpretability and utility in clinical settings, aiding doctors in summarizing patient symptoms and providing preliminary diagnostic insights, thereby facilitating further diagnosis and having a substantial role in clinical psychological consultations.

\begin{table}[h]
\caption{Results of ablation studies}
\centering
\setlength{\tabcolsep}{2.5pt}{ 
\begin{tabular}{cccccc}
\hlineB{4}
\textbf{Model} & \textbf{BLEU-2} & \textbf{ROUGE-L} & \textbf{METEOR} & \textbf{DIST-2} \\
\hline
ChatGLM3 +POST* & \textbf{45.28\%} & \textbf{0.56} & \textbf{0.5794} & 0.10 \\
w/o Info & 44.43\% & 0.54 & 0.5708 & 0.16 \\
w/o Stage & 43.57\% & 0.53 & 0.5681 & 0.16 \\
w/o Sum & 44.25\% & 0.54 & 0.5722 & 0.16 \\
w/o Next & 36.24\% & 0.45 & 0.5107 & \textbf{0.18} \\
\hlineB{4}
\end{tabular}}
\label{tab:ablation-results}
\end{table}

\section{Conclusions}

 In this paper, we incorporate Psychological State Tracking (POST) within LLM to guide response generation for doctors during depression diagnosis consultations. In particular, the state is defined based on Albert Ellis’ ABC Model in psychology, which illuminates a profound connection between patients' information changes and doctors' strategic planning. 
Extensive experiments show that the integration of psychological state tracking significantly enhances the performance of LLMs to generate responses in depression-diagnosis-oriented chat. Furthermore, our approach also provides explicit interpretations for using appropriate strategies in different situations to collect information or comfort patients for depression diagnosis.
In the future, we will explore more specialized and fine-grained state tracking methods and incorporate patient personalized information to guide diagnosis-oriented chat.

\subsubsection{Acknowledgement} This research is funded by the National Nature Science Foundation of China (No. 62477010 and No.62307028), the Natural Science Foundation of Shanghai (No. 23ZR1441800), Shanghai Science and Technology Innovation Action Plan (No. 24YF2710100 and No.23YF1426100) and Shanghai Special Project to Promote High-quality Industrial Development (No. 2024-GZL-RGZN-02008).

%
%
%
\bibliographystyle{splncs04}
\bibliography{post}

\end{document}